\documentstyle[twoside,fleqn,espcrc2]{article}
\newcommand{\AmS}{{\protect\the\textfont2
  A\kern-.1667em\lower.5ex\hbox{M}\kern-.125emS}}
\newcommand{\Tr}{{\rm Tr}}

\newcommand{\plaq}{{\rm plaq}}
\newcommand{\third}{\mbox{$\frac{1}{3}$} }

\title{Heavy Quark Physics
\thanks{Review presented at {\em Lattice 92}, Amsterdam, Sept. 15--19, 1992.}
}

\author{Paul B. Mackenzie \\
\medskip
Theoretical Physics Group\\
Fermi National Accelerator Laboratory\\
P. O. Box 500\\Batavia, IL 60510 USA}

\begin{document}

\begin{abstract}

\end{abstract}

\maketitle

\section{INTRODUCTION}
Present and future
lattice calculations involving $b$ and $c$ quarks include some of the
most important applications of lattice gauge theory to standard model
physics.
These include the heavy meson decay constants, the $B \overline{B}$
mixing amplitude, and various semileptonic decay amplitudes,
which are all crucial in extracting CKM angles from experimental data.
They also include the extraction of $\alpha_s$ from the charmonium
and bottomonium spectra.

Bound states of heavy quarks and antiquarks (quarkonia) have another
crucial role to play in the development of lattice gauge theory:
they provide systems in which the estimation of the errors inherent in
current lattice calculations can be done in a more reliable and robust
way than is possible for the light hadrons.
The reason is that the quarks in these systems are relatively nonrelativistic.
Coulomb gauge wave functions calculated on the lattice may be used
to aid in the estimation of finite volume and finite lattice spacing errors,
and of the effects of quenching.
We have a much better idea of what to expect in lattice calculations
of these systems since potential models may be used to obtain the leading
behavior in $v^2/c^2$.\cite{quigg}

Chris Sachrajda and I will split the subject of heavy quark physics in
these proceedings.  His review \cite{Sachrajda}
will concentrate on the part of the subject
which involves the weak interactions.
Mine will concentrate on the part which does not.

\section{LATTICE FORMULATIONS OF HEAVY QUARKS}\label{form}

When the lattice spacing $a$ is smaller than the Compton wave length of the
quark $1/m$, the standard relativistic action of Wilson may be used.
Cutoff effects may be removed by taking the cutoff $1/a$ to infinity.
The bare lattice action may also be viewed as an effective field theory
of QCD at the cutoff scale.
Cutoff effects in an effective field theory are
 removed by adding higher dimension  interactions to the bare Lagrangian
while keeping the cutoff fixed.
To remove the effects of the cutoff to a finite order in $a$,
a finite number of interactions may be added to the bare lattice
Lagrangian.
When $a \Lambda_{QCD} \ll 1$, perturbation theory may be used to
calculate the required coefficients of the new operators.\cite{Symanzik}
  The ability
to remove cutoff effects perturbatively will probably be spoiled eventually,
perhaps at a small power of $a$ due to effects presently not understood,
almost certainly at a relatively large power of $a$ due to instantons.

The dynamical scales in bound states are small compared to the fermion
mass in QED and in QCD for the $c$ and $b$ quarks.
It is often advantageous in these systems to formulate the
field theory nonrelativistically as an expansion in
$1/m$ \cite{ThackerLepage,EichtenHL,Feinberg,Caswell},
keeping the cutoff at or below $m$.
The nonleading terms in nonrelativistic expansions have dimension higher than
four.  Loop corrections in these effective field theories
diverge if the cutoff is removed.
Cutoff effects  must be removed by adding higher
dimension interactions to the Lagrangian or by raising the cutoff to
the new physics scale ($m$),
 switching to the
relativistic, renormalizable version of the theory, and then taking
the cutoff to infinity.

When the kinetic energy of the heavy quark is small compared to the
typical interaction energies (as it is in bound states containing a single
heavy quark), the kinetic energy may be
treated as a perturbation.  In this {\em static
approximation} \cite{Feinberg,EichtenHL}, the lowest order fermion
action is just
\begin{equation}
{\cal L}_{\mbox{static}}=\phi^* i D_t \phi,
\end{equation}
and the unperturbed quark propagator is just the timelike Wilson
line.
In the general case, including quarkonia,
the lowest order potential {\em and } kinetic terms of the Lagrangian
of {\em Nonrelativistic QCD} (NRQCD) \cite{Caswell,ThackerLepage}
(the terms on the first
line of Equation \ref{LNRQCD}) must be included in the unperturbed
Lagrangian.
\begin{eqnarray}\label{LNRQCD}
{\cal L}_{\mbox{\tiny NRQCD}}=\phi^* \{i D_t &+&
    \frac{{\bf D}^2}{2 m} \nonumber \\
&+& \frac{g}{2 m} {\bf \sigma \cdot B} \}   \phi \nonumber \\ &+& \cdots
\end{eqnarray}
Higher order terms in $\frac{1}{m}$ may be added as perturbations.

In processes such as the semileptonic decay of heavy-light mesons, in
which one heavy quark decays into another lighter (but still heavy)
quark with a high velocity relative to the first, it is possible, and
useful, to formulate the static approximation and nonrelativistic QCD
as expansions in the small internal quark momentum around some large,
external meson momentum.\cite{Isgur} Lattice implementations of this
idea have been proposed \cite{Mandula} and are reviewed by
Sachrajda.\cite{Sachrajda}

\subsection{The improvement program for Nonrelativistic QCD}\label{NRQCD}

In a recent paper \cite{fiveauthors}, Lepage et al. have
systematically examined the improvement program for NRQCD with the
goal of reducing systematic sources of error from all sources to under
10\%.  This program involved the following elements:

1) Since NRQCD has been  formulated as a nonrenormalizable effective field
theory, cut-off effects are removed not by taking the cut-off to
infinity, but by adding additional operators to the bare Lagrangian
(the dots in Eq. \ref{LNRQCD}).  The infinite number of possible
operators must be ordered according to expected size of their effects
on the physics.  For heavy-light systems, the operator ordering is
simply an expansion in $1/m$, that is, in the dimensions of the
operators.  For heavy-heavy systems like quarkonia, the expansion is
complicated by the presence of large quark velocities which do not
fall to zero with the quark mass.  Operators with the same dimension
(such as $\frac{{\bf D}^2}{2 m}$ and $ \frac{g}{2 m} {\bf \sigma \cdot
B}$ ) are suppressed in their effects on the physics by different
powers of $v$ (by $v^2$ and $v^4$, respectively, in this case).

2) Once the operators required for a given accuracy have been
established, their coefficients must be determined by requiring that
the NRQCD Lagrangian reproduce the Green's functions of ordinary QCD
to this accuracy.

3) Discrete forms of the required operators must then be defined.  As
with light quark actions, finite $a$ errors must be estimated. If
necessary, correction operators \cite{Symanzik} must be added to the
action.

4) The coefficients of the operators are modified by quantum effects.
Many corrections have been calculated in mean field
theory.\cite{fiveauthors}  The corrections for the quark energy
shift, mass renormalization, and wave function renormalization have
been calculated in full one-loop perturbation theory.\cite{BAT}
Deviations between the mean field and one loop results are rather
small, from 0--10\%.

The result is a systematic correction program in $v$, $a$, and
$\alpha_s$.  The correction operators in $v$ and $a$ may be included
directly in the simulation action, or evaluated as perturbations
using lattice or potential model wave functions.

\subsection{A New Action for Four Component Fermions}\label{general}

Because coefficients of higher terms in the NRQCD Lagrangian such as
$\frac{{\bf D}^2}{2 m}$ are explicit functions of $1/m$, the quantum
corrections described in 4) above are also explicit functions of
$1/m$.  These begin to diverge as $m a$ is reduced below a value of
order one, making the nonrelativistic expansion impractical.  The Wilson
action likewise has been thought to have finite lattice spacing errors
of order $m a$ which blow up as $m a$ is raised above one.  Since the
masses of the $b$ and $c$ quarks are such that $ma$ is often $O(1)$ at
current lattice spacings, calculations of such crucially interesting
quantities as the heavy meson decay constants $f_B$ and $f_D$ have
often involved awkward interpolations between results in the static
approximation and results using Wilson fermions through a region where
neither approximation is well behaved.\cite{Sachrajda} While such an
approach is probably workable, it would clearly be desirable to have a
method for lattice fermions which did not begin to break down right in the
region of interest.

To approach such a method, we consider a lattice version of ${\cal
L}_{\mbox{\tiny NRQCD}}$ with a few minor modifications.
  Like Wilson fermions ($\psi$), the fermions of
NRQCD contain four components per site: a two-component quark field
($\phi$) and a two-component antiquark field ($\chi$).
The  bare mass is conventionally
omitted in NRQCD calculations, but we are free to leave it in the
theory.  The usual Dirac coupling between quarks and
antiquarks is absent (having been transformed into higher derivative
interactions by the Foldy-Wouthhuysen transformation), but we may add
back a sufficiently suppressed amount of this interaction without
spoiling the theory.  We thus consider the following Lagrangian:
\begin{eqnarray}
{\cal L}&=& \phi^* (c_1 \ \Delta_t^- + m_0 - \frac{c_2}{2} \sum_{i}
\Delta^+_i\Delta^-_i)\phi_n \nonumber \\ \nonumber &+& c_3 \ \phi^*
\sum_{i} \sigma_i \Delta_i \chi_n \\ &+& \chi^* (- c_1 \ \Delta_t^+ +
m_0 - \frac{c_2}{2} \sum_{i} \Delta^+_i\Delta^-_i)\chi_n \nonumber \\
&+& c_3 \ \chi^* \sum_{i} \sigma_i \Delta_i \phi_n.
\end{eqnarray}
When $c_1=1$ (times a correction factor when $m a \gg 1$),
$c_2=\frac{1}{m}$, and $c_3$ is negligible, it is a good Lagrangin for
NRQCD.  The point of writing the NRQCD Lagrangian in this particular
form is that the action becomes precisely the standard Wilson action
with the choice of parameters $c_1=c_2=c_3=1.$ It is thus possible to
adjust the parameters in such a way that as $m_0$ is reduced, instead
of blowing up, the theory turns smoothly into the Wilson theory.

It is illuminating to expand the equation for Wilson propagators
nonrelativistically when the mass is large.  After normalizing the
fields by $\frac{1}{\sqrt{1-6\kappa}}$ \cite{lunorm} (not
$\frac{1}{\sqrt{2\kappa}}$ as is conventional) one may obtain
\begin{eqnarray}\label{SE}
\lefteqn{\delta_{{\bf n} 0} = [{\cal -E + M}  + (1-U^\dagger_{{\bf n},0} ) }\\
  && -\frac{1}{2}\left( \frac{1}{ m_0 } +
\frac{1}{(1+m_0)\ (2+m_0) } \right)  \sum_{i} ( \Delta_i )^2
]\phi_{\bf n}\nonumber ,
\end{eqnarray}
where $\cal E$ is the energy eigenvalue obtained from the transfer
matrix and ${\cal M=E}_{p^2=0} = \ln(1+m_0)$.  This is a lattice
Schr\"odinger equation not unlike the one obtained from NRQCD, but it
has some unusual features.  Most important, the two ``masses'' in the
equation, ${\cal M}= \ln(1+m_0)$ and $\frac{1}{M} = \frac{1}{ m_0 } +
\frac{1}{(1+m_0)\ (2+m_0) }$, are completely different.
If $\cal M$ is used to fix the fermion mass when $a m \gg 1$, the
dynamically more important mass condition $\partial {\cal E}/ \partial
p^2 = \frac{1}{2m}$ will be completely incorrect.

Kronfeld showed in his talk at this conference \cite{KronMack} that
the two masses can be put back into agreement with the use of the
action
\begin{eqnarray} \label{newaction}
S &=& \sum_{n} [ -\bar{\psi}_{n} \psi_{n} \nonumber\\ &+&\kappa_t
\bar{\psi}_{n}(1-\gamma_0) U_{n,0}
\psi_{n+\hat{0}} + h. c.		\nonumber\\
&+&\kappa_s\sum_{i}\bar{\psi}_{n}(1-\gamma_i) U_{n,i}
\psi_{n+\hat{i}} + h. c.  ]
\end{eqnarray}
Thus, it seems that an action closely related to the Wilson action is
a member of the class of actions suitable for NRQCD.  One can go even
further.  In NRQCD and in the static approximation, $\cal M$ plays no
dynamical role. It can be ignored, and is conventionally thrown away.
This suggests that the standard Wilson action itself can be used when
$a m > 1$ as long as $\cal M$ is ignored and $\partial {\cal E}/
\partial p^2 = \frac{1}{2m}$ is used to fix the quark mass, as is done
in NRQCD.

This proposal is obviously correct in free field, where we can
calculate the behavior of quark propagators exactly to see that the
proposed interpretation makes sense.  It is certainly correct in mean
field theory, too.  Mean field improvement of these fermions, as of
Wilson fermions, is simply the absorption of a ``mean link'' $u_0$
(see Appendix \ref{PT}) into an effective $\tilde{\kappa} \equiv u_0
\kappa$ and then proceeding as with free field theory.  (A plausible
estimate of the mean link in this context is probably $u_0 \sim
1/8\kappa_c$.)  It remains to be shown whether the theory is somehow
spoiled by renormalization.

Perturbatively, Green functions must be expanded in $p^2$ and
$\alpha_s$.  Each term in the expansion is an explicit function of the
quark mass, since the theory must be solved exactly in $m a$. (The is
also the case for the loop corrections of NRQCD.\cite{BAT} If these
functions become singular or badly behaved in some way, the theory
could conceivably break down.  The one loop perturbative corrections
contain all of the ugliest features of Wilson and NRQCD perturbation
theory simultaneously, and have only been begun.  There is, however,
one numerical calculation by El-Khadra \cite{cnorm} indicating that
nothing too surprising occurs.  The one-loop correction to the local
current normalization for Wilson fermions with the naive normalization
is \cite{Zhang}
\begin{equation}   \label{currentnorm1}
\langle \psi | V_4^{loc} | \psi \rangle =
 \frac{1}{2 \kappa (1-0.17 g^2) }.
\end{equation}
The correct normalization with mean field improvement is
\begin{equation}   \label{currentnorm}
\langle \psi | V_4^{loc} | \psi \rangle =
 \frac{1}{(1-\frac{6\kappa}{8\kappa_c})(1-0.06g^2) }.
\end{equation}
The remaining perturbative correction, $0.06g^2$, becomes an explicit
(so far uncalculated) function of $m$ (or $\kappa$) in the new
formalism which must not become singular if the theory is to make
sense.
\begin{figure}[ht]
\vspace{7.1truecm}
\caption{Normalization of the local vector current as a function of $\kappa$.}
\label{fig:currentnorm}
\end{figure}

Fig. \ref{fig:currentnorm} shows Eqs. \ref{currentnorm1} (upper curve)
and
\ref{currentnorm} (lower curve) along with a numerical calculation of the
quantity at two values of $\kappa$ ($16^3x32$ lattice, $\beta=5.9$).
It can be seen that for this
quantity, not only is the unknown function of $m$ not singular, it is
approximately equal to 1.

Putting the new action on a secure footing will ultimately require: 1)
determination of the bare parameters of the action with mean field
theory and full perturbation theory, 2) nonperturbative tests of the
perturbative results, and 3) phenomenological tests of the resulting
action in calculations of well understood physical quantities.  Not
much of this program has yet been accomplished.  However, as argued
above, at large values of $m a$, the new action (and even the Wilson
action suitably reinterpreted) can be viewed simply as unusual members
of the general class of lattice actions proposed by Lepage and
collaborators for NRQCD.  Quite a bit is now known about the action
for NRQCD.  The discussion in points 1) and 2) of Sec. \ref{NRQCD} on
operator classification is valid for any method for treating heavy
quarks, including this one.  The fact that the mean field corrections
discussed in 4) reproduce the mass-dependent one loop corrections very
well is encouraging.

Care will clearly be required in formulating normalization conditions
which capture the most important physics in both the relativistic and
nonrelativistic regions.  (Identifying $\partial {\cal E}/ \partial
p^2$ rather than $\cal M$ as the fundamental mass condition is example
number one of these.)

\section{PHENOMENOLOGY OF THE $J/\psi$ AND $\Upsilon$ SYSTEMS}\label{onia}
Like all phenomenological lattice calculations, calculations of the
properties of heavy quark systems serve a variety of purposes.
Quantities which are well understood experimentally, but which are
very sensitive to lattice approximations are good tests of lattice
methods (Sec. \ref{HFS}).  Quantities for which the lattice
approximations are well understood may be used to extract information
about the standard model (Sec. \ref{1p1s}).  A further purpose for
lattice calculations is the delineation of the limits and the reasons
for the successes of earlier models of hadrons (Sec. \ref{comppot}).

I will discuss calculations in the $\psi$ and $\Upsilon$ systems by
Davies, Lepage, and Thacker \cite{Davies} using NRQCD, and
calculations in the $\psi$ system by the Fermilab group using Wilson
fermions reinterpreted as described in Sec. \ref{general}
\cite{El-Khadra+al,PBM,AXE,AXE2,fermicharm}, and by UKQCD using Wilson
fermions \cite{Allton}.  (See also \cite{NRQCDUKQCD}.)  Both groups
studying the $J/\psi$ system used the $O(a)$ correction term
\begin{equation} \label{eq:WOSH}
\delta {\cal L}=
-i g \frac{c}{2} \overline{\psi} \Sigma_{\mu\nu} F_{\mu\nu} \psi
\end{equation}
of Sheikholeslami and Wohlert~\cite{WOSH}.  UKQCD used the tree level
coefficient $c=1$.  The Fermilab group used a mean field improved
coefficient $c=1.4$ (see Appendix \ref{PT}).

\subsection{1S--1P Splitting}\label{1p1s}
An excellent determination of the lattice spacing in physical units is
provided by the spin averaged splitting between the lowest angular
momentum ($l=0$ and $l=1$) levels of the $\psi$ and $\Upsilon$
systems.  (In the charm system, for example, $M_{h_c} - (3 M_\psi
+M_{\eta_c})/4 = 458.6 \pm 0.4$ MeV.)  The values of the lattice
spacing obtained from this splitting do not differ dramatically from
those obtained from other quantities, such as the $\rho$ mass
\cite{Weingarten} or the string tension \cite{Bali2}.  It is the
possibility of making improved uncertainty estimates that makes this
an important way of determining the lattice spacing.  In quarkonia,
error estimates may often be made in several ways: by brute force (e.
g., by repeating the calculation several lattice spacings), by
phenomenological arguments, and by direct calculation of correction
terms.  Since determination of the lattice spacing is one of the key
components of the determination of the strong coupling constant from
low energy physics, it is important that these uncertainty estimates
be made rock solid.

Preliminary results for this mass splitting were reported last year by
the Fermilab group \cite{El-Khadra+al,PBM} and by Davies, Lepage, and
Thacker \cite{Davies}.  This year, El-Khadra \cite{AXE2} reported
further work done to check the corrections and error estimates given
in Ref. \cite{El-Khadra+al}.  In \cite{El-Khadra+al}, uncertainties
due to an incorrectly known quark mass and to $O(a)$ errors arising
from an imperfectly determined coefficient $c$ in the $O(a)$
correction to the Lagrangian (Eq. \ref{eq:WOSH}) were taken to be less
than 1\% and omitted from the table of errors on the basis of the
phenomenological arguments.  (The splitting is expected to be
insensitive to small errors in the definition of the quark mass since
it is almost identical in the $\psi$ and $\Upsilon$ systems. Likewise,
in quark models the contribution of the ${\bf \sigma}\cdot{\bf B}$
interaction, which dominates $\delta {\cal L}$ nonrelativistically, to
the spin averaged splitting is zero.)  These arguments were checked
this year by repeat calculation at several values of the parameters
and found to be correct within statistical errors.

The $O(a^2)$ errors were argued to be small because repeat
calculations at $\beta=5.7$, 5.7, and 6.1 yielded almost the same
result for $\alpha$(5 GeV) (see next section).  An attempt was made to
correct for the small variation observed by extrapolating to zero
lattice spacing in $a^2$.  This extrapolation is not completely
satisfactory, since the small $a$ functional form is a messy
combination of $O(a^2)$ errors and perturbative logarithms.  The way
to improve this result, which has not yet be done, is to follow the
example the NRQCD group \cite{Davies} and evaluate directly the
contributions of the known correction operators to the splitting,
thereby eventually obtaining zero measurable dependence on the lattice
spacing.  This group evaluated the correction of the operators
perturbatively using the wave functions of the Richardson
\cite{Richardsonpot} potential model.
For the $\Upsilon$ at $\beta=6.0$, for
example, they obtained the rather small corrections shown in Table
\ref{table:NRQCD}.
  Wave functions directly
calculated by lattice gauge theory could also be used, eliminating the
need for potential models. They are easy to calculate to high statistical
accuracy.  Fig. \ref{fig:WF} shows the Coulomb gauge wave function
of the $J/\psi$ meson calculated on a $24^4$ lattice at $\beta=6.1$.
Statisitical errors are negligible at small separations.

\begin{figure}
\vspace{7.4truecm}
\caption{The wave function of the $J/\psi$ meson.}
\label{fig:WF}
\end{figure}

Similarly, the estimate of the finite volume correction needs to be
bolstered by calculating the meson Coulomb gauge wave functions on the
lattice, and then calculating the overlap integral of the wave
function with its periodically reflected image.

\begin{table} \centering
\begin{tabular}{c|c|c}
Term		&$\Delta M(1P-1S)$	& \%	\\
\hline
$O(v^2)$		& -11 MeV	& -2\%	\\
$O(a_t)$		& 13 MeV	& 3\%	\\
$O(a^2_x)$		& -12 MeV	& -3\%	\\ $\delta
S_{gluon}$	& -24 MeV	& -5\%	\\
\hline
Total			& -33 MeV	& -7\%	\\
\end{tabular}
\caption{ NRQCD corrections to the 1P--1S splitting in the $\Upsilon$
system.  Finite lattice spacing corrections are for $\beta=6.0$.
}\label{table:NRQCD}
\end{table}

\subsection{Determination of $\alpha_s$ from the 1S--1P Splitting}

The most recent determinations of $\alpha_s$ from the charmonium and
bottomonium spectra using NRQCD \cite{Davies} and modified
Wilson fermions \cite{fermicharm} have error bars bracketing
the region $\alpha_s = 0.103-0.114$ GeV.
They are somewhat below, but consistent with the world average given in
the review of QCD in the 1992 particle data book.  They are inconsistent
with the most recent LEP determinations, which are around 0.120 and
above.\cite{Ellis}

The determination of $\alpha_s$ from the 1S--1P splitting currently
consists of three separate elements: the determination of the lattice
spacing, the determination of a physical coupling constant at a scale
measured in lattice spacing units, and, for the time being, a
correction for the absence of light quarks.  As discussed in the
previous section, the uncertainties in the determination of $\alpha_s$
arising from the determination of the lattice spacing seem to be in
good shape right now, and the path is clear to making them very solid.

\subsubsection{Determination of the coupling constant.}

To determine the running coupling constant, one would like to combine
the determination of the lattice spacing discussed above with a
nonperturbative calculation of a physically defined coupling constant,
for example defined from the static quark potential at a given, fixed
momentum transfer like 5 or 10 GeV.

Since the largest cutoff momenta for the existing 1P--1S splitting
calculations was $\pi/a \approx 7.5$ GeV, it was not possible in the
existing calculations to determine the continuum limit of a physical
coupling defined at short distances.

In Ref. \cite{El-Khadra+al} a mean field improved perturbative
relation, (Eq. \ref{MFI} in Appendix \ref{PT}) was used to obtain a
renormalized coupling from the bare lattice coupling.  This relation
was tested over the past year on short distance Wilson
loops.\cite{LM2} It did well, but not perfectly: the loops calculated
by Monte Carlo were systematically a few per cent high.  This suggests
that Eq. \ref{MFI} fixes most but not all of the pathological relation
between the bare lattice coupling constant and physical couplings, and
that it is better to obtain physical couplings from short distance
quantities calculated nonperturbatively.  Using short distance Wilson
loops for this purpose (for example, via Eq. \ref{logplaq}) raises the
values of the renormalized couplings by a few per cent over those
reported in Ref. \cite{El-Khadra+al}.

It remains to be determined how much couplings defined from continuum
quantities differ from those defined from short distance quantities
like the log of the plaquette.  There is some reason to expect that
this difference is small.  The second order corrections to the short
distance lattice static potential are within a few per cent of the
continuum corrections.\cite{Heller} Likewise, Creutz ratios of Wilson
loops up to six by six are quite well behaved when expanded with a
coupling defined from Eq.
\ref{logplaq}.\cite{LM2}

The Monte Carlo calculation of the static quark potential at a
separation of one lattice spacing agrees to very high accuracy with
perturbation using the coupling of Eq. \ref{logplaq}.  (See Sec.
\ref{pertpot}.)  Therefore, a coupling constant obtained from the very
short distance static potential will give results almost identical to
those using Eq. \ref{logplaq}.  A phenomenological method for
estimating the continuum coupling constant defined by the potential
using short distance data has been proposed by Michael.\cite{Michael1}
It has been used by UKQCD \cite{Michael2} and by Bali and
Schilling.\cite{Bali2} It yields results which are quite close to
those obtained with Eqs. \ref{MFI} and \ref{logplaq}, and therefore to
those which would be obtained directly from short distance lattice
potential itself.

A program to calculate explicitly a continuum coupling constant using
finite size scaling has been proposed by L\"uscher et
al.\cite{Luscher-etc} To select the particular coupling to focus on,
they propose the criteria that it: 1) be defined nonperturbatively, 2)
be calculable in perturbation theory, 3) be calculable in Monte Carlo
simulations, and 4) have small, controllable lattice artifacts.  These
lead them to propose the response of the QCD vacuum to a constant
background color-electric field to define the coupling constant.  The
more phenomenological choice of the static potential has poorer
signal-to-noise properties in Monte Carlo calculations approaching the
continuum limit, and (they tell us) more difficult higher order
perturbation theory.  An SU(2) calculation has been completed, which
yields results similar to those obtained with Eq. \ref{MFI}.  An SU(3)
calculation is in progress.

\subsubsection{Correction for the effects of sea quarks.} \label{seaquarks}
This is the greatest source of uncertainty in the results quoted above.
This correction is the most phenomenological, and has the greatest
likelihood of having a problem.  Over the next few years, it will be
removed by direct inclusion of the effects of sea quarks.

The attempt to estimate the effect that the absence of sea quarks has
on this result is based on three assumptions.  They are, in order of
decreasing rigor:
\begin{itemize}
\item When certain physics quantities are used to tune bare parameters in
the quenched approximation, the most important terms in the effective
Lagrangian at the dominant energy scale for those quantities are given
correctly.  The effective action at other energy scales including the
scale of the lattice cutoff will be somewhat incorrect.  In
particular, if the effective coupling constant at the physics scale
approximates that of the real world, the effective coupling at the
short distance cutoff will be a bit small.
\item The most important term in the effective action for quarkonia
is the static potential.  The phenomenological success of potential
models indicates that this assumption may be valid to around 25\%.  It
is this assumption, which is certainly not valid for the light
hadrons, that leads us to dare to try to make this correction for the
charmonium system when we would not try it for the light hadrons.
\item The effects of light quarks on the static potential
may be estimated by fitting charmonium data with a QCD based potential
model such as the Richardson potential once with the correct, $n_f=3$,
$\beta$ function in the potential and again with the quenched,
$n_f=0$, $\beta$ function.
\end{itemize}

The final assumption is certainly a good one at short distances, which
are responsible for most of the difference in the evolution of the
coupling from the middle distance charmonium physics scale down to the
lattice cutoff scale.  It is also reasonable at the less relevant
large distance scale, since the lattice quenched string tension and
the string tension of Regge phenomenology are comparable.  If,
however, light quarks have a much greater effect on the potential in
middle distances than they seem to at large and at small distpp ances,
the assumption would fail.

The naive expectation for the size of the correction is
$$ \frac{\beta^{n_f=0}_0- \beta^{n_f=3}_0}{ \beta^{n_f=3}_0 }\sim
20\%.  $$
In Ref. \cite{El-Khadra+al}, a perturbative calculation was used to bound
the plausible size of the correction.
This year the estimated correction was checked \cite{fermicharm}
by fitting the charmonium spectrum with a potential twice: once using
a potential with the correct $\beta$ function and once using
a potential with the quenched $\beta$ function.  (See
Sec. \ref{HFS})
The result was compatible with the one in Ref. \cite{El-Khadra+al}.
This, however, is not so much an independent check of the previous estimate
as another
 quantification of the assumption that sea quarks have no more dramatic
effects on the potential at middle distances than they seem to at large
and small distances.

\subsubsection{Future prospects for determining $\alpha_s$}

Errors in the determination of the lattice spacing are already in good
shape.

 The accuracy in the determination of the coupling constant needs
further examination, but the calculations of Ref. \cite{LM2} suggests
that the accuracy to be expected of lattice perturbation theory is
greater than that expected of QCD perturbation theory in hadronic
phenomenology.  In Ref. \cite{LM2}, discrepancies of about $\alpha^2$
were typically observed in comparisons of first order perturbation
theory with Monte Carlo calculations, and discrepancies of about
$\alpha^3$ in comparisons of second order perturbation theory.  This
amounts to only 3-4\% for calculations at the lattice cutoff at
moderate $\beta$'s.  In contrast, in QCD phenomenology, an accuracy of
10\% is often taken to be optimistic.  One difference may be that the
lattice calculations of most interest are often quadratically
divergent integrals dominated by momenta of the order of the
relatively well-defined lattice cutoff.  They thus differ from
calculations of unruly hadrons in collision, which insist on
interacting on a wide range of momentum scales, piling up large
logarithms from a nasty variety of sources.

The aspect of the current determination of $\alpha_s$ which makes it
no better than any other existing determination is the use of
potential model arguments to estimate the effects of the absence of
sea quarks.  This is quite analogous to, for example, the
phenomenological treatment of higher twist and fragmentation effects
in determinations of $\alpha_s$ in deep inelastic scattering.  All of
the existing determinations have some phenomenological assumptions
built into them.  The difference is that the potential model estimate
of quenched corrections will certainly be eliminated by brute computer
force (if not by the use of more intelligent methods) over the next
few years, resulting in a determination far more accurate than any of
the existing ones.

\subsection{Hyperfine Splitting and Leptonic Width}\label{HFS}

These two quantities are very straightforward to calculate on the
lattice, and are good phenomenological tests of how well we understand
the parameters of the quark action.  The hyperfine splitting $\Delta m
(J/\psi - \eta_c) $ and leptonic decay amplitude $V_\psi \equiv
m^2_\psi/f_\psi $ have been calculated in the $J/\psi$ system by the
Fermilab group and by UKQCD.  They have been calculated in the
$\Upsilon$ system by Davies, Lepage, and Thacker.\cite{Davies}

The potential model formula for the hyperfine splitting is
\cite{quigg}
\begin{equation}
\Delta m (J/\psi - \eta_c)  =
	\frac{32 \pi \alpha_s(m_c)}{9 m_c^2} |\Psi(0)|^2 \;\;.
\end{equation}
It arises from a coupling of the spins of the quarks to transverse
gluons.  It therefore should be extremely sensitive to the value of
the correction coefficient $c$.  (It is not clear {\em a priori}
whether to expect strong sensitivity to the quark mass, since $
|\Psi(0)|^2$ should rise with the quark mass.)

The leptonic width to leading order is
\begin{equation}
\Gamma_{ee}  =
	\frac{16 \pi \alpha^2 e^2_c }{ m_c^2} |\Psi(0)|^2 \;\;.
\end{equation}
Nonrelativistically, the leptonic decay amplitude is therefore simply
the wave function at the origin, $\Psi(0)$, properly normalized.  This
quantity should be quite sensitive to the mass of the quark.

Before comparing existing lattice results with experiment, we need to
estimate the accuracy to expect in the quenched approximation.  Both
quantities are proportional to $ |\Psi(0)|^2 $, the probability for
the quarks to be at the same point (within one Compton wave length,
say) and so are obviously short distance quantities.  With lattice
parameters tuned to obtain the correct 1P--1S splitting, the coupling
constant and potential at short distances will be too weak.  An
analysis like the one referred to in Sec. \ref{seaquarks} yields
\begin{equation}
\frac{\alpha_s^{(0)} (m_c)}{\alpha_s^{(3)} (m_c)} = 0.81 \pm 0.06 \;\;.
\end{equation}

The incorrect weakness of the quenched potential at short distance may
also yield a weakened wave function at the origin.  El-Khadra has
checked for this effect \cite{AXE2,fermicharm} using the Richardson
potential
\begin{equation}
V(q^2) = C_F \frac{4\pi}{\beta^{(n_f)}_0} \frac{1}{q^2
\ln{(1+q^2/\Lambda^2)}} \;\;,
\end{equation}
where  $\beta_0^{(n_f)} = 11 - 2 n_f /3$ .
Fitting the charmonium spectrum with once
 with $n_f=3$ in the $\beta$ function parameter,  and again with
$n_f=0$,  she found that, indeed, The ratio of the wave functions
at the origin was
\begin{equation} \label{eq:wfcor}
\frac{\Psi^{(0)}(0)}{\Psi^{(3)}(0)} = 0.86 \;\;.
\end{equation}
This reduces our expectation for the hyperfine splitting from the
experimental result $\Delta m (J/\psi - \eta_c)^{\rm exp} = 117.3$ MeV
to around
\begin{equation}  \label{eq:hypquen}
\Delta m (J/\psi - \eta_c)^{\rm quenched} \simeq 70 \; {\rm MeV} \;\;.
\end{equation}
These considerations reduce our expectations for
 leptonic matrix element by a smaller amount,  from the experimental result
$V_{\psi}^{\rm exp} = 0.509$ GeV$^{3/2}$ to
\begin{equation} \label{eq:Vquen}
V_{\psi}^{\rm quenched} = 0.438 \; {\rm GeV}^{3/2} \;\;.
\end{equation}

\begin{figure}
\vspace{8.3 truecm}
\caption{$\Delta m (J/\psi - \eta_c)  $ calculated by UKQCD (diamond)
and Fermilab (squares) compared with the physical result (upper star)
and an estimate of the quenched corrected result (lower star).   }
\label{fig:HFS}
\end{figure}

The results of UKQCD and Fermilab for the hyperfine splitting are
shown in Fig. \ref{fig:HFS}.  UKQCD set the value of the quark mass to
obtain the expected energy eigenvalue in the transfer matrix.  In
light of the arguments on the interpretation of Wilson fermions at large
quark masses
in Sec. \ref{general}, the Fermilab group took
this as unreliable and attempted to fix the quark mass by demanding
that the leptonic width be correct.  The UKQCD result is slightly
below the result expected on the basis of the quenched correction.
This is consistent with the mean field expectation that quantum
corrections boost the required value for the coefficient of the
correction term.  Their results are, however, much closer to the
physical answer than earlier Wilson fermion calculations with no
$O(a)$ correction.\cite{Boch} The Fermilab results are slightly above
the quenched expectation, but perhaps not very significantly in light
of the uncertainties in the quenched correction and the statistical
errors.

\section{THE STATIC QUARK POTENTIAL}\label{pot}
High accuracy results for the static quark potential were reported
this year by Bali and Schilling \cite{Bali1,Bali2} and by UKQCD
\cite{Michael2}.  Fig. \ref{fig:potential} shows the potential
calculated by Bali and Schilling in the quenched approximation on a
$32^4$ lattice at $\beta=6.4$.  The solid line is the fit to a Coulomb
plus linear potential
\begin{equation}
V(R) = V_0 -0.277(28)/R +0.0151(5) R,
\end{equation}
which fits quite well for $R>2\sqrt{2}.$ Comparison of results on
$16^4$ with results on $32^4$ lattice indicated that finite volume
results are small.  A plot of results from $\beta=6.0$, 6.2, and 6.4
with physical units set by the string tension indicates good scaling
behavior.

\begin{figure}
\vspace{5.3truecm}
\caption{The heavy quark potential, calculated on the lattice in the quenched
approximation.}
\label{fig:potential}
\end{figure}

\subsection{Short distance behavior.}  \label{pertpot}
At such a large $\beta$ we should expect the short distance part of
the potential to agree very well with perturbation theory, and this is
the case.  I checked the value of $V(1)$ given in Ref. \cite{Bali2}
against perturbative results for $32^4$ lattices supplied by Urs
Heller \cite{Urs}.  Using the ``measured'' coupling constant defined
by Eq. \ref{logplaq}, perturbation theory agreed with the Monte Carlo
data to within about 1\%, perhaps fortuitously accurate, but still
impressive.  (This incidentally illustrates that the potential is a
natural candidate on the lattice as well as in the continuum to define
improved coupling constants.  The coupling of Eq. \ref{logplaq} was
suggested mostly because the plaquette is easy to measure and
universally available.)

Since perturbation theory agrees so well with the Monte Carlo
calculation of the potential, and since perturbation theory implies a
coupling constant rising with increasing $R$, it would be interesting
to attempt to fit the data with an asymptotically free Coulomb plus
linear potential.  The size of the fit Coulomb term ($0.277(28)/R$) is
quite close to the subleading long distance behavior of the potential
($\pi/(12R)=0.262/R$).  However these two similarly-sized effects have
nothing to do with each other, and we are not guaranteed that, for
example, the perturbative Coulomb term does not rise above 0.28 before
the potential settles back down to its asymptotic form.  A fit with an
asymptotically free Coulomb plus linear potential, for example a
modified Richardson potential \cite{Richardson}, might help to start
exploring the extent to which the data support or rule out such
speculation.

It is easy to convince yourself with a ruler that values of the string
tension obtained by the fit are completely plausible.  However, if the
changeover from the perturbative Coulomb potential to the
nonperturbative long distance $1/R$ term is more complicated than we
hope, a larger than expected middle distance Coulomb term could be
contaminating the obtained string tensions more than is obvious from
the current analysis.

\subsection{Long distance behavior.}

The good scaling of the potential when the physical scale is set by
the string tension has already been mentioned.  Good {\em asymptotic}
scaling of the string tension in terms of a physical coupling such as
$\Lambda_{\overline{MS}}$ is also observed in the new data.  (Good
means to perhaps 20\%.)  Folklore to the contrary was based on the the
search for scaling in terms of the bare lattice coupling constant. The
bare coupling has a highly pathological, but reasonably
well-understood relationship to well-behaved physical coupling
constants.  It was pointed out long ago \cite{Karsch,Karsch2} that
decent scaling is observed in terms of an effective coupling constant
defined from the plaquette.  It was emphasized in Ref. \cite{LM2} that
such coupling constants are simply very close relations of the
familiar physical coupling constants such as $\alpha_V$ and
$\alpha_{\overline{MS}}$ of perturbative QCD.

\begin{figure}
\vspace{7.5truecm}
\caption[fig:]{$\sqrt{\sigma}/\Lambda_{\overline{MS}}$ as a function of the
lattice spacing.  The upper curve was obtained from the bare coupling
constant.  The lower curve was obtained from an effective coupling
constant. }
\label{fig:siglam}
\end{figure}

Fig. \ref{fig:siglam} (from Ref. \cite{Karsch2}) shows new and old
data for $\sqrt{\sigma}/\Lambda_{\overline{MS}}$ plotted as a function
of $a \Lambda_{\overline{MS}}$.  The upper curve
was obtained via the bare coupling constant.  The much better behaved
lower curve was obtained via the effective coupling
\begin{eqnarray}
\alpha_{eff}=
\frac{3( 1-\frac{1}{3}Tr (U))} {4\pi}.
\end{eqnarray}
(Ref. \cite{LM2} advocates Eq. \ref{logplaq}, the logarithm of $Tr
(U)$, for this purpose on the grounds that logarithms of Wilson loops
have better perturbative behavior than the loops themselves.)  Only
about 20\% deviation from asymptotic scaling is observed over the
range of the data.  Part of that deviation is certainly perturbative,
since the use of another reasonable perturbative scheme, Eq.
\ref{logplaq}, changes the amount of deviation to 10\%.\cite{LM2} The
fact that the ratio of the deconfinement temperature to the square
root of the string tension scales better than
$\sqrt{\sigma}/\Lambda_{\overline{MS}}$ is another indication that the
deviation is more likely to be connected to the determination of the
coupling constant than to the string tension.  Because the short
distance behavior is a mixture of perturbative logarithms and $O(a^2)$
errors, the extrapolation in $a$ is not completely satisfactory and it
is important to sort the origin(s) of the discrepancy: perturbation
theory, $O(a^2)$ errors, or measurement errors.  However, the downward
trend seems clear and the estimate
\begin{eqnarray}
\sqrt{\sigma}/\Lambda_{\overline{MS}} = 1.75 \pm 15\%
\end{eqnarray}
seems reasonable.

\subsection{Comparison with potential models.} \label{comppot}

\begin{figure}
\vspace{6truecm}
\caption{The lattice quenched
heavy quark potential (top curve) and the potentials of the Cornell
model and the Richardson model.}
\label{fig:phenpot}
\end{figure}
One useful task of first-principles calculations is to support or
destroy earlier phenomenologies.  For example, it would be nice to be
able to understand if there is a reason that nonrelativistic quark
models for the light hadrons work unreasonably well.  It is more
straightforward to put the success of potential models of heavy quark
systems on a rigorous footing using lattice methods.  These systems
are nonrelativistic and it is not surprising that a nonrelativistic
treatment yields rather accurate results.

In Fig. \ref{fig:phenpot} the potential obtained by Bali and Schilling
is compared with the potentials of Eichten et al. \cite{Eetal} and
Richarson in the region $0.1 \ \mbox{fm} < R < 1.0 \ \mbox{fm} $.  The
string tensions of the lattice and the phenomenological potentials are
similar, but the Coulomb term required by phenomenology is about 1.8
times as large as that yielded by the quenched lattice, seemingly a
large discrepancy.  The phenomenological potential is very well known
in this region between 0.1 and 1.0 fm.  Fits to the spectra of the
charmonium and bottomonium systems with a wide variety of plausible
and implausible functional forms yield potentials which differ by only
a few per cent in this region.  On the other hand, the quenched
lattice potentials are also rather convincing, especially at short
distance, so what accounts for the difference?  First, we expect the
quenched Coulomb coupling to be a bit smaller than the true QCD
coupling constant at short distances because of the incorrect $\beta$
function of the quenched approximation.  (See Sec. \ref{seaquarks}.)
This effect is in the right direction and is expected to be of order
$$ \frac{\beta^{n_f=0}_0- \beta^{n_f=3}_0}{ \beta^{n_f=3}_0 }\sim
20\%.  $$ Second, the phenomenological potentials clearly parameterize
some of the effects of higher order relativistic corrections.  These
are roughly expected to be of order $v^2/c^2 \sim 25\%$ for
charmonium.  Some of these clearly have the effect of strengthening
the attraction of the quarks, but a complete analysis of the
spin-independent relativistic corrections in potential models does not
exist.\cite{Gromes} A combination of these two effects could thus
easily explain as much as 1.5 out of the discrepancy of 1.8.  A
preliminary conclusion: there is an interesting puzzle in this
discrepancy, but no cause for alarm.

\section{SUMMARY}
{\bf Static Potential.}

\begin{itemize}
\item At short distances, the potential agrees with perturbation theory to
a few per cent.
\item The string tension exhibits two loop asymptotic scaling to an
accuracy of 20\%.
$\sqrt{\sigma}/\Lambda_{\overline{MS}} $ is in the range
1.55--1.95.
\end{itemize}

{\bf $\psi$ and $\Upsilon$ systems.}

\begin{itemize}
\item The hyperfine splitting and leptonic widths provide good phenomenological
tests of lattice methods.
\item The spin averaged 1P--1S splitting provides a very good determination
of the lattice spacing in physical units.  Combined with a lattice
determination of the renormalized coupling, it gives a determination
of the strong coupling constant which at present is of comparable
accuracy to that of conventional determinations.  When the effects of
sea quarks are properly included, its accuracy will be much better
than any current determination.

\end{itemize}

 {\bf Technical developments.}

\begin{itemize}
\item Lattice perturbation theory works very well when renormalized
coupling constants are used.
\item  Minor changes to the actions of Wilson and of NRQCD may make it
possible do calculations with a unified formalism at any value of the
quark mass, as long as the three momentum is small.  This will imply a
reinterpretation of calculations with Wilson fermions at large quark
mass.

\end{itemize}

\section*{ACKNOWLEDGEMENTS}
I would like to thank Peter Lepage, Estia Eichten, Aida El-Khadra,
Andreas Kronfeld, and Chris Sachrajda for helpful discussions.

	Fermilab is operated by 	Universities Research
Association, Inc. under contract with the U.S.  	Department of
Energy.

\appendix
\section{NOTATION} \label{note}
We use the forward, backward, and symmetric finite difference
operators
\begin{eqnarray}
\Delta^+_\mu \psi_n & \equiv & U_{n,\mu} \psi_{n+\hat{\mu}} -\psi_n,	\\
\Delta^-_\mu \psi_n & \equiv &
\psi_n-U^\dagger_{n-\hat{\mu},\mu}\psi_{n-\hat{\mu}},	\\
\Delta_\mu \psi_n   & \equiv & \frac{ \Delta^+_\mu
+\Delta^-_\mu}{2} \psi_n.
\end{eqnarray}
The analogous continuum covariant derivative is denoted
\begin{eqnarray}
D_\mu
\end{eqnarray}

The standard relation between the bare mass $m_0$ and the hopping
parameter $\kappa$ is
\begin{equation}
m_0 \equiv \frac{1}{2\kappa } -4.
\end{equation}

When considering nonrelativistic fermions, we decompose the
four-component Dirac field as two two-component fields
\begin{equation}  \label{psiphichi}
\psi = \left(\begin{array}{c}
	\phi \\ 	\chi
\end{array}  \right),
\end{equation}
and take as our representation of the Euclidean gamma matrices
\begin{equation}
\gamma_0 = \left(\begin{array}{cc}
	1 & 0 \\ 	0 & -1
\end{array}  \right),\ \ \ \ \ \ \ \
\gamma_i = \left(\begin{array}{cc}
	0 		& \sigma_i \\ 	\sigma_i 	& 0
\end{array}  \right).
\end{equation}

\section{RESULTS FROM LATTICE PERTURBATION THEORY \hfill}\label{PT}
This section summarizes results from Ref. \cite{LM2} which have been
used in the text.

\subsection{A sequence of improved coupling constants}

In Ref. \cite{LM1} (1990) it was argued that lattice perturbation
series are much more convergent and agree better with Monte Carlo data
if they are expressed in terms of a physical running coupling
evaluated at a carefully chosen scale.  A good one is $\alpha_V$, the
one defined by the static quark potential:
\begin{equation}
  \frac{1}{\alpha_V(q)} = \frac{1}{\alpha_{lat}} + \beta_0
\ln(\frac{\pi}{a q}) - 4.702.
\end{equation}
The arguments were analogous to those leading from the $MS$ to the
$\overline{MS}$ scheme in dimensionally regularized QCD.

In Ref. \cite{El-Khadra+al} (1991) it was noted that the bulk of the
correction coefficient in the previous equation is accounted for by a
simple mean field argument.  The coupling constant is enhanced at one
loop by a coupling to the expectation value of the plaquette induced
by the higher order terms in the Wilson action.  Higher order
analogues of this one loop effect certainly exist.  This suggests that
an effective coupling constant which incorporates a Monte Carlo
calculation of the plaquette expectation value, such as
\begin{equation}\label{MFI}
  \frac{1}{\alpha_V(q)} = \frac{\langle Tr U_p \rangle}{ \alpha_{lat}}
+ \beta_0 \ln(\frac{\pi}{a q}) -0.513,
\end{equation}
may yield improved accuracy.

Over the past year (1992) we have tested this assumption by
calculating a variety of short distance quantities using the mean
field improved coupling constant and by Monte Carlo.\cite{LM2} We
found that, while using Eq. \ref{MFI} significantly improved agreement
between perturbation theory and Monte Carlo, the Monte Carlo results
tended to be systematically slightly higher then the perturbative
results.  This suggests that a coupling defined directly from any of
the Monte Carlo calculated quantities would yield improved predictions
for the others.  A particularly simple one is the coupling defined
from the log of the plaquette:
\begin{eqnarray} \label{logplaq}
\frac{1}{\alpha_V(3.41/a)}&=&
\frac{1}{\alpha_{\it eff}}-1.19   \nonumber \\
& \equiv & -\frac{4\pi}{3 \ln(\frac{1}{3}Tr (U))} - 1.19
\end{eqnarray}
(The scale of the running coupling arises from an estimate of the
typical momenta of gluons in the calculation of the logarithm of the
plaquette.\cite{LM2})

\subsection{Mean field improvement of operators}

The mean field argument leading to Eq. \ref{MFI} may be summarized as
follows.  The naive classical relation between the lattice and
continuum gauge fields \begin{equation} \label{naive} U_\mu(x) \equiv
e^{iagA_\mu(x)} \:\to\: 1 + iagA_\mu(x).  \end{equation} is spoiled by
tadpoles arising from the exponential form of the lattice
representation of the gauge fields.  Quantum fluctuations do not lead
to an average link field close to 1.00000 as implied by Eq.
\ref{naive}, but to something more like \begin{equation}
\label{improved} U_\mu(x) \to u_0\,(1+iagA_\mu(x)), \end{equation}
where $u_0$, a number less than one, represents the mean value of the
link.  In a smooth gauge, the Monte Carlo link expectation value can
be used as an estimate of $u_0$.  A simple, gauge-invariant definition
is \begin{equation} u_0 \equiv \langle\third \Tr U_\plaq
\rangle^{1/4}.  \end{equation} Other definitions based on $\kappa_c$
or the static quark self-energy may be used to fine tune mean field
predictions in particular situations.

If naive definitions such as Eq. \ref{naive} are used to relate
lattice and continuum operators, large corrections will appear in
quantum corrections.  Much better behavior of loop corrections is
obtained by taking $ U_\mu(x)/ u_0 $ as the lattice approximation to
the continuum field.  This implies that the lattice action
\begin{equation} \tilde{S}_{\rm gluon} = \sum \frac{1}{\tilde{g}^2
u_0^4} \Tr (U_{\plaq} + {\rm h.c.}).  \end{equation} will approximate
closely the desired continuum behavior.  This is the usual lattice
action if we identify \begin{eqnarray} \tilde{g}^2 &=& g^2_{lat}/u_0^4
= g^2_{lat}/\langle \mbox{$\frac{1}{3}$}\Tr(U_{\rm plaq})\rangle.
\end{eqnarray} The perturbative result, Eq. \ref{MFI}, explicitly
verifies that $ \tilde{g}^2$ is a closer approximation to a standard
continuum expansion parameter than $ g^2_{lat}$ is.\cite{Parisi}

The same considerations lead to the result that
\begin{equation} \label{kappa_eff}
\tilde{\kappa}\equiv\kappa u_0
 \end{equation} produces a more continuum-like bare mass ($
\tilde{m}=\frac{1}{2 \tilde{\kappa}}-4 $) and smaller quantum
corrections in operator renormalizations than does $\kappa$.

Mean field arguments may be used both to estimate perturbative
predictions when the perturbative predictions are unknown, and also to
improve known predictions.  Just as we did with the coupling constant
in Eq. \ref{MFI}, we can improve perturbative predictions for
operators involving quark fields by substituting the Monte Carlo
calculation of $u_0$ in Eq. \ref{kappa_eff} and including in the
perturbative prediction only that part remaining after the absorption
of $u_0$ into $\kappa$.  A possible fine tuning in this case is to
obtain $u_0$ from $\kappa_c$ rather than from the plaquette.  This was
the procedure used in obtaining Eq. \ref{currentnorm}.

The cloverleaf approximations to $F_{\mu\nu}$ used in the
Wohlert-Sheikholeslami $O(a)$ correction to the Wilson action
\cite{WOSH} and in the magnetic spin coupling of NRQCD
\cite{fiveauthors} contain four links each.  The quark wave function
normalization contains one link.  We therefore expect the naive
coefficient of this operator to undergo quantum corrections of roughly
a factor of $u_0^{-3}$, or about $1+0.25 g^2$ if we use the plaquette
to estimate $u_0$.  This is in agreement with an unpublished thesis
calculation of Wohlert, $ 1 + 0.27 g^2$.\cite{WO} Using the plaquette
calculated by Monte Carlo to estimate the correction term yields a
factor of $\langle \mbox{$\frac{1}{3}$}\Tr(U_{\rm plaq})\rangle^{-3/4}
\sim 1.4 - 1.5 $ for $\beta$ around 6.0.


\begin{thebibliography}{9}


\bibitem{quigg} For a review on heavy quarkonia see for example:
 	W. Kwong, J. L. Rosner, C. Quigg, Ann. Rev. Nucl. Part. Sci.
	{ 37} (1987) 325:
	the importance of quarkonia to lattice gauge theory has been
	emphasized by Lepage, \cite{Lepage}.
\bibitem{Lepage} G. P. Lepage, in {\em Lattice 91}, M. Fukugita et al.,
	editors, Nuc. Phys. B (Proc. Suppl.) 26 (1992) 45.
\bibitem{Sachrajda} C. T. Sachrajda, in these proceedings.
\bibitem{Symanzik} K. Symanzik, Nucl. Phys. B226 (1983) 187, 205.
\bibitem{ThackerLepage} G. P. Lepage and B. A. Thacker,
  in {\em Field Theory on the Lattice,} proceedings of the
International Symposium, Seillac France, 1987, edited by A. Billoire
et al.  [Nuc. Phys. B (Proc. Suppl.) 4 (1988) 199]; and Phys. Rev. D
43 (1991) 196.
\bibitem{EichtenHL} E. Eichten,
  in {\em Field Theory on the Lattice,} proceedings of the
International Symposium, Seillac France, 1987, edited by A. Billoire
et al.  [Nuc. Phys. B (Proc. Suppl.)accirac 4 (1988) 170].
\bibitem{Feinberg} E. Eichten and F. L. Feinberg, Phys. Rev. Lett. 43 (1979)
	1205.
\bibitem{Caswell} W. E. Caswell and G. P. Lepage, Phys. Lett. 167B (1986) 437.
\bibitem{Isgur} N. Isgur and M. B. Wise, Phys. Lett. B232 (1989) 113.
\bibitem{Mandula} J. Mandula and M. Ogilvie, Phys. Rev. D45 (1992) 2183.
\bibitem{fiveauthors} G.P.~Lepage, L.~Magnea, C.~Nakhleh, U.~Mag\-nea, and
	K.~Hornbostel, Cornell preprint CLNS 92/1136, to be published
in 	Phys. Rev. D. See also \cite{Lepage}
\bibitem{BAT} C. T. H. Davies and B. A. Thacker, Phys. Rev. D45 (1992) 915.
\bibitem{lunorm} This is the normalization used in
	M. L\"uscher, Commun. Math. Phys. 54, (1977) 283.
\bibitem{KronMack} A. S. Kronfeld and P. B. Mackenzie, in preparation;
	A. S. Kronfeld, contribution to  these proceedings.
\bibitem{cnorm} Which she has seen fit to include in the contribution
	of A. S. Kronfeld to these proceedings.
\bibitem{Zhang} G. Martinelli and Y.-C. Zhang, Phys. Lett. 123B (1983) 433.
\bibitem{Davies} C. T. H. Davies, G. P. Lepage, and B. A. Thacker, (1991,1992)
	to be 	published.
\bibitem{El-Khadra+al} A. X. El-Khadra, G. Hockney, A. S. Kronfeld, and P. B.
	Mackenzie, Phys. Rev. Lett. 69 (1992) 729.
\bibitem{PBM} P. B. Mackenzie, in {\em Lattice 91}, M. Fukugita et al.,
	editors, Nuc. Phys. B (Proc. Suppl.) 26 (1992) 369.
\bibitem{AXE} A. X. El-Khadra, in {\em Lattice 91}, M. Fukugita et al.,
	editors, Nuc. Phys. B (Proc. Suppl.) 26 (1992) 372.
\bibitem{AXE2}  A. X. El-Khadra, contribution to these proceedings.
\bibitem{fermicharm} A. X. El-Khadra, G. Hockney, A. S. Kronfeld, and P. B.
	Mackenzie, to be published.
\bibitem{Allton} C. R. Allton et al., UKQCD collaboration, Southampton
	preprint SHEP 91/92-27 (1992).
\bibitem{NRQCDUKQCD} While this review was being completed, a new preprint
	by UKQCD appeared on the network:  S. M. Catterall et al.,
	DAMTP-92-70.
\bibitem{WOSH} B.~Sheikholeslami and R.~Wohlert, Nucl. Phys. B259 (1985) 572.
\bibitem{Weingarten} F. Butler, H. Chen, A. Vaccarino, J. Sexton,
	and D. Weingarten, contribution to these proceedings.
\bibitem{Bali2}
	G. S. Bali and K. Schilling, Wuppertal preprint WUB 92-29 (1992).
\bibitem{Richardsonpot} J. Richardson, Phys. Lett. 82B (1979) 272.
\bibitem{Ellis} R. K. Ellis, review given at DPF 92,
	Fermilab, November 10-14, 1992.
\bibitem{LM2} G. P. Lepage and P. B. Mackenzie, Fermilab preprint 91/355-T
	(Revised) (1992).
\bibitem{Heller} U. Heller and F. Karsch, Nuc. Phys. B251 [FS13] (1985) 254.
\bibitem{Michael1} C. Michael, Phys. Lett. B283 (1992) 103.
\bibitem{Michael2} S. P. Booth et al., UKQCD Collaboration, Liverpool
	preprint LTH 285 (1992).
\bibitem{Luscher-etc} M.~L\"uscher, R.~Sommer, U.~Wolff and 
	 P.~Weisz,	CERN preprint
	CERN-TH \linebreak  6566/92, and references therein; talk by
	M. L\"uscher in these proceedings.
\bibitem{Boch} M. Bochicchio et al., Nuc. Phys. B372 (1992) 403.
\bibitem{Bali1}
	G. S. Bali and K. Schilling, Wuppertal preprint WUB 92-02 (1992).
\bibitem{Urs} I thank Urs Heller for these results, 
	obtained with
	programs written for Ref. \cite{Heller}
\bibitem{Richardson} The potential used in Ref. \cite{Richardsonpot},
	 modified to make the linear
	coefficient independent of the Coulomb coefficient may be a
	reasonable one to try for this purpose.
\bibitem{Karsch} F.~Karsch and R.~Petronzio, Phys. Lett. 139B (1984) 403.
\bibitem{Karsch2}
	J.~Fingberg, U.~Heller and F.~Karsch, Bielefeld preprint BI-TP~92-26.
\bibitem{Eetal}
	 E. Eichten, K. Gottfried, T. Kinoshita, K. D. Lane, and
	T.-M. Yan, Phys. Rev. D17 (1978) 3090.
\bibitem{Gromes} As opposed to the analysis of the spin-dependent relativistic
	corrections by E. J. Eichten and F. Feinberg, Phys. Rev. Lett. 43
	(1979) 1205, and Phys. Rev. D 23 (1981) 2724; W. Buchmuller,
	Phys. Lett. 112B (1982) 479;
	and D. Gromes, Z. Phys. C 26 (1984) 401.
\bibitem{LM1} G. P. Lepage and P. B. Mackenzie, in {\em Lattice 90}, U. M.
	Heller et al., editors, Nuc. Phys. B (Proc. Suppl.) 20 (1991) 173.
\bibitem{Parisi} Another version of this argument is given by
G. Parisi, in {\em High Energy Physics--1980}, proceedings
of the XX International Conference, Madison, Wisconsin, L. Durand and L. G.
Pondrom, editors, American Institute of Physics (1981).
\bibitem{WO} R.~Wohlert, Ph.D. Thesis, unpublished
	DESY preprint 87/069.


\end{thebibliography}
\end{document}